\documentclass[twocolumn]{article}
\usepackage[a4paper, total={180mm, 243mm}]{geometry}
\usepackage[utf8]{inputenc}
\usepackage{amssymb}
\usepackage{lipsum}
\usepackage{graphicx}
\usepackage{siunitx}
\usepackage[version=4]{mhchem} 
\usepackage{titling}
\usepackage{circledsteps}   
\usepackage{subcaption}

\usepackage[switch]{lineno} 
\modulolinenumbers[5]

\DeclareCaptionLabelSeparator{custom}{\textbar}
\DeclareCaptionFormat{custom}{\textsf{\textbf{#1 #2} \small #3}}
\newcommand\authormark[1]{\textsuperscript{#1}}

\graphicspath{{./figures}} 

\makeatletter
\newcommand{\showfontsize}{\f@size{} pt}
\makeatother

\usepackage[
    backend=biber,
    style=nature,
    url=false,
    isbn=false,
    eprint=false,
    date=year
]{biblatex}
\newcommand{\reffig}[2]{Fig.~\ref{#1}\textbf{\textsf{#2}}}

\definecolor{Myblue}{cmyk}{0.71,0.53,0,0.12}
\definecolor{Myblue2}{rgb}{0,0,0.75}
\newcommand{\highlightchanges}[1]{#1}

\title{Synthetic-reflection self-injection-locked microcombs}

\author{Alexander~E.~Ulanov,\authormark{1}
        Thibault Wildi,\authormark{1}
        Nikolay~G.~Pavlov,\authormark{2} \\
        John~D.~Jost,\authormark{2}
        Maxim Karpov,\authormark{2}
        Tobias Herr\authormark{1,3,*}}

\date{%
    \small $^1$Deutsches Elektronen-Synchrotron DESY, Notkestr. 85, 22607 Hamburg, Germany \\
    \small $^2$Enlightra Sarl, Rue de Lausanne 64, 1020 Renens, Switzerland\\
    \small $^3$Physics Department, Universität Hamburg UHH, Luruper Chaussee 149, 22607 Hamburg, Germany\\
    \small $^*$tobias.herr@desy.de
}

\begin{document}

\maketitle

\textbf{
Laser-driven microresonators have enabled chip-integrated light sources with unique properties, including the self-organized formation of ultrashort soliton pulses and frequency combs (microcombs). While poised to impact major photonic applications, such as spectroscopy, sensing and optical data processing, microcombs still necessitate complex scientific equipment to achieve and maintain suitable single-pulse operation. Here, to address this challenge, we demonstrate microresonators with programmable synthetic reflection providing an injection-feedback to the driving laser. When designed appropriately, synthetic reflection enables \highlightchanges{deterministic} access to self-injection-locked microcombs operating exclusively in the single-soliton regime. These results provide a route to easily-operable microcombs for portable sensors, autonomous navigation, or extreme-bandwidth data processing. \highlightchanges{The novel concept of synthetic reflection may also be generalized to other integrated photonic systems.}
}

\begin{refsection}
Laser-driven microresonators provide access to nonlinear optical phenomena, already with low-power continuous-wave excitation \cite{vahala:2003}. 
Leveraging efficient nonlinear frequency conversion, they have enabled novel sources of coherent laser radiation across broad spectral span \cite{delhaye:2007, sayson:2019}. 
Soliton microcombs \cite{Herr2014, kippenberg:2018, diddams:2020} are an important representative of such sources, providing frequency comb spectra of mutually coherent laser lines, based on self-organized dissipative Kerr solitons (DKSs) in resonators with anomalous group velocity dispersion (GVD) \cite{leo:2010}.
Such DKS microcombs can be integrated on photonic chips \cite{brasch:2016, gaeta:2019} and have demonstrated their disruptive potential in many emerging and ground-breaking applications, e.g. high-throughput optical data transmission \cite{Marin-Palomo2017} reaching Pbit-per-second data rates \cite{jorgensen:2022}, ultrafast laser ranging \cite{trocha:2018, suh:2018}, precision astronomy in support of exo-planet searches \cite{suh:2019, Obrzud2019}, high-acquisition rate dual-comb spectroscopy \cite{suh:2016}, ultra-low noise microwave photonics \cite{liang:2015, lucas:2020}, photonic computing and all-optical neural networks \cite{feldmann_parallel_2021, xu:2021, bai:2023}. 

To leverage microcomb technology in out-of-lab applications, it is critical to reliably access the DKS regime and ideally \textit{single}-DKS operation \cite{guo:2017}, ensuring well-defined temporal and spectral characteristics. 
A critical challenge for microcombs \highlightchanges{arises from the need to stabilize} the detuning $\Delta \omega_0 = \omega_0 - \omega_\mathrm{p}$ of the pump laser $\omega_\mathrm{p}$ with respect to the pumped resonance $\omega_0$. While this is common to all resonant approaches, it is particularly challenging during DKS initiation, when thermo-optic effects can cause a rapid ($\sim\mu$s) change in resonance frequency \cite{Herr2014}. To overcome this challenge, a number of methods have been developed, involving rapid laser actuation \cite{Herr2014, brasch:2016}
, auxiliary lasers \cite{zhang_sub-milliwatt-level_2019} and/or auxiliary resonances \cite{li:2017, weng_dual-mode_2022}, laser modulation \cite{wildi:2019}, additional nonlinearities \cite{he:2019, rowley:2022, bai:2021} or, pulsed driving \cite{obrzud:2017}. \highlightchanges{These methods are successfully used in research.}

\highlightchanges{An attractive approach that can stabilize the laser detuning for DKS operation is self-injection locking (SIL). It relies on a feedback wave created through backscattering in the microresonator that effectively locks the laser frequency to the microresonator resonance \cite{vasilev:1996, liang:2010, Kondratiev2017, jin:2021}. SIL has been utilized for DKS generation in bulk whispering-gallery mode resonators \cite{liang:2015, Pavlov2018}, as well as, in highly-integrated photonic chip-based systems \cite{stern:2018, raja_electrically_2019, Shen2020, Voloshin2021, xiang:2021}}. \highlightchanges{In these SIL-based DKS sources, the feedback wave is based on Rayleigh backscattering from random fabrication imperfections or material defects in the microresonator \cite{gorodetsky:2000}. However, critically relying on random imperfections is incompatible with scaling of microcomb technology into large volume application. It is also at odd with the intense efforts towards reducing backscattering through improved fabrication processes.}


\begin{figure}[ht!]
  \centering
  \includegraphics[width=\columnwidth]{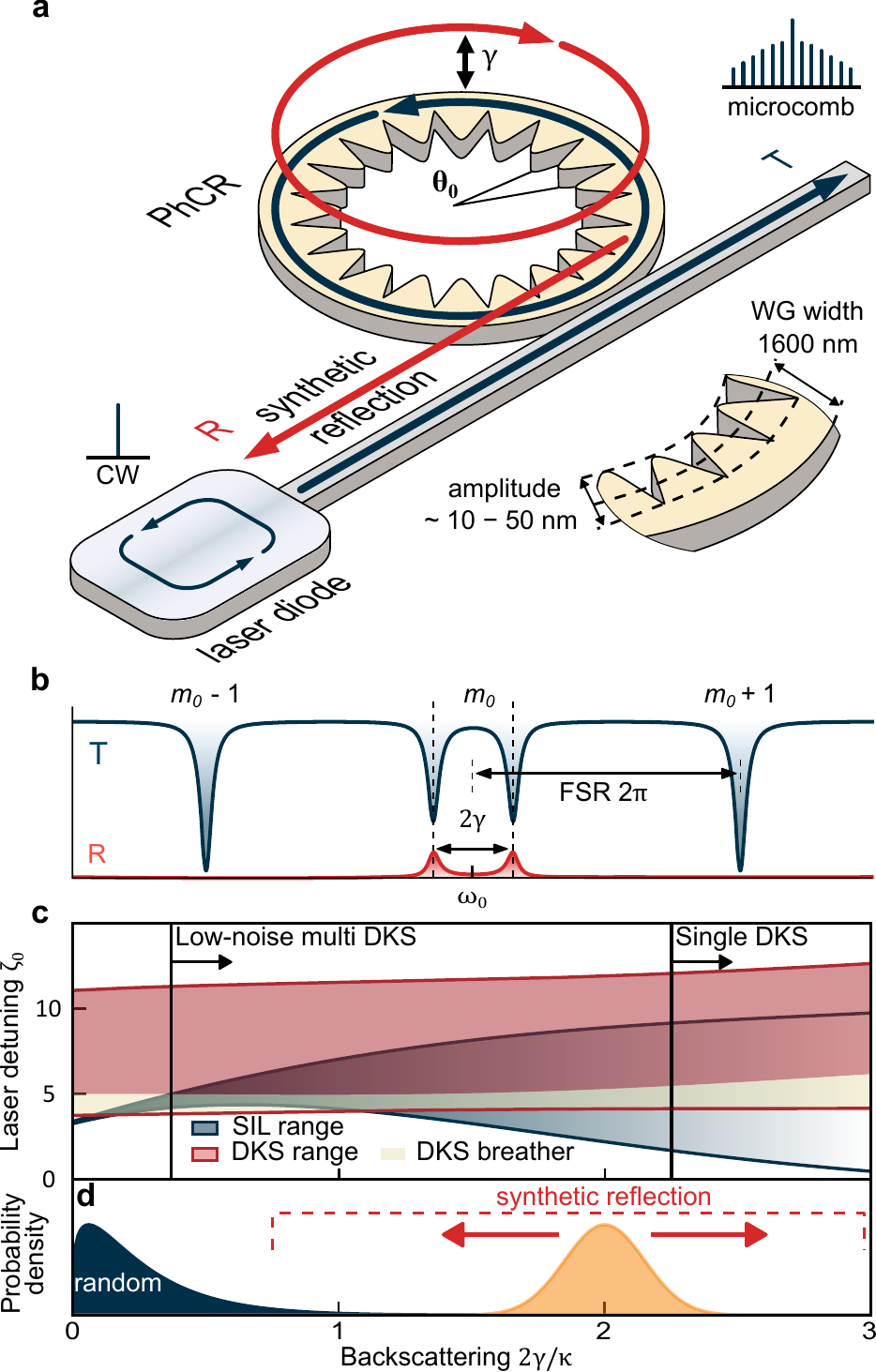}
  \caption{
      \highlightchanges{\textbf{Self-injection locking with synthetic reflection}.
        \textbf{a}, Integrated photonic crystal ring-microresonator (PhCR) with a periodic corrugation (angular period $\theta_0$), which induces coupling at a rate $\gamma$ between forward and backward-propagating waves for a mode $m_0 = \pi / \theta_0$. In addition to a transmission signal (T), this leads to a well-defined synthetic resonant reflection (R), which can be programmed for self-injection locking with a laser diode driving the system. The magnified view, indicates the typical dimensions.
        \textbf{b}, Indicative transmission and reflection spectrum for the resonance with mode number $m_0$ and two adjacent resonances $m_{0\pm1}$, separated by $\pm1$ free-spectral range (FSR). For $\gamma\neq 0$, the lineshape at mode number $m_0$ exhibits a split lineshape (frequency splitting $2\gamma$), and shows non-zero resonant reflection.
        \textbf{c}, Comparison of (nonlinear) SIL and DKS existence ranges (cf. main text) in units of normalized detuning $\zeta_0$ (cf. Methods) as function of backscattering for a critically coupled microresonator with total linewidth $\kappa/2\pi = 120$ MHz, dispersion $D_2/2\pi = 8$ MHz, driven with a normalized pump power of $f^2 = 9$. 
        \textbf{d}, Illustration of the probability to find a certain strength of backscattering in a conventional microresonator (blue) or in a PhCR with synthetic reflection (orange); the PhCR with synthetic reflection allows for much larger and tailored backscattering by design.}
    }
  \label{fig:intro}
\end{figure}

\highlightchanges{
Here, we introduce \textit{synthetic reflection} as a novel method for self-injection locked microcombs. In contrast to previous demonstrations, this method is independent from random backscattering and permits the generation of tailored backreflection spectrum without disturbing the dispersion profile or noticeably decreasing the quality-factor ($Q$). This is achieved deliberately in photonic crystal ring resonators (PhCR) (Figure~\ref{fig:intro}a) \cite{arbabi_realization_2011}, which have recently received growing attention in integrated nonlinear photonics \cite{
Yu2021, lu_high-q_2022, black_optical-parametric_2022, yang:2022, lucas:2022}. We show that synthetically created  backreflection can provide robust access to DKS states by increasing the overlap between laser detunings where DKS can exist and those detunings that can be accessed via SIL. In addition, we show that robust access to SIL-based DKS can be combined with recent results of spontaneous single-DKS generation in PhCRs, avoiding non-solitonic states \cite{Yu2021}. Significantly, we derive analytic criteria, that permit designing the synthetic reflection to ensure exclusive operation in the single-DKS regime. These results provide a route to easily-operable microcombs for out-of-lab applications.}


\subsection*{Results}

To gain independence from random imperfections, we use PhCRs that enable synthetic reflection by design. The reflection is controlled by periodic nano-patterned corrugations of the ring-resonators' inner walls. The angular corrugation period is  $\theta_0=2\pi/(2m_0)$, where $m_0$ is the angular (azimuthal) mode number, for which a deliberate coupling between forward and backward propagating waves with a coupling rate $\gamma$ is induced (see \reffig{fig:intro}{a}). Besides inducing the desired synthetic reflection, the coupling leads to mode hybridization resulting in a split resonance lineshape (frequency splitting $2\gamma$) in both transmission and reflection (see \reffig{fig:intro}{b}). Here, we only consider the lower frequency hybrid mode for pumping, as it corresponds to strong (spectrally local) anomalous dispersion, which prevents high-noise comb states \cite{herr_universal_2012}. For choosing $\gamma$ we balance multiple criteria, as we detail below:

First, a strong reflection can significantly extend the range of normalized detunings $\zeta_0 = 2\Delta \omega_0 / \kappa$ ($\kappa$ is the microresonator linewidth) accessible via SIL (SIL range) in a nonlinear microresonator. This is crucial, as it permits robust access to detunings where DKS can exist (DKS existence range). This is exemplified in \reffig{fig:intro}{c}, where the SIL range according to the theory by Voloshin et al. \cite{Voloshin2021} is shown along with the numerically computed DKS and breathing DKS existence ranges (obtained through integration of the coupled mode equations cf. Methods). Note, that in a resonator with a shifted pump mode \cite{helgason_power-efficient_2022}, the existence range of DKS deviates strongly from that known from resonators without a shifted pump mode \cite{godey:2014} and can currently only be obtained numerically (\reffig{fig:intro}{c}). In conventional resonators the normalized forward-backward coupling is usually small $2\gamma/\kappa < 1$ \highlightchanges{(\reffig{fig:intro}{d}) and the intersection between SIL and DKS ranges does not exist or is limited; synthetic reflection can reliably provide access to larger backscattering.}

Second, while advantageous for an extended SIL range, stronger forward-backward coupling $\gamma$ will also result in an increased threshold \highlightchanges{at which modulation instability (MI) occurs}. Without MI, DKS cannot form inside the resonator (without external stimuli such as triggering pulses~\cite{leo:2010}). The threshold power is different from that in a conventional ring resonator \highlightchanges{\cite{lobanov:2020}} and its derivation critically requires consideration of the backward wave. For strong forward-backward coupling ($2\gamma/\kappa>1$), the following approximation is derived (cf. Supplementary information - SI) \highlightchanges{for the threshold pump power}
\begin{equation}
    f_\mathrm{th}^2 = 4\frac{\gamma}{\kappa} + \frac{\kappa}{\gamma}
    \label{eq:f2th}
\end{equation}
\highlightchanges{which is normalized as detailed in the Methods.} The value of $f_\mathrm{th}^2$ must not exceed the available pump power $f^2$. If the MI threshold is reached at a detuning within the DKS existence range, then the MI state is only transient and DKS  can form spontaneously \highlightchanges{as recently observed in PhCRs \cite{Yu2021} and over-moded resonators \cite{tan:2020} owing to the DKS attractor \cite{Herr2014, barashenkov:1996, balakireva:2018}.}

Third, \highlightchanges{by controlling the MI state that precedes the DKS, the deterministic creation of a single-DKS state can be achieved. This is the case when the first modulation instability sidebands are separated from the pump laser by 1~FSR, corresponding to a modulated waveform with a single intensity maximum. Such behavior has been observed in PhCRs driven by an external continuous-wave laser \cite{Yu2021}. Here, assuming symmetric power in forward and backward pump mode, we estimate analytically the condition for exclusive single-DKS formation (cf. SI)}
\begin{equation}
    \frac{\gamma}{\kappa} >  \frac{f^2}{8}
    \label{eq:gamma1DKS}
\end{equation}
\highlightchanges{Note, that due to the small power-asymmetry between forward and backward modes, the actual required value will be slightly higher.}

\begin{figure}[ht!]
  \centering
  \includegraphics[width=\columnwidth]{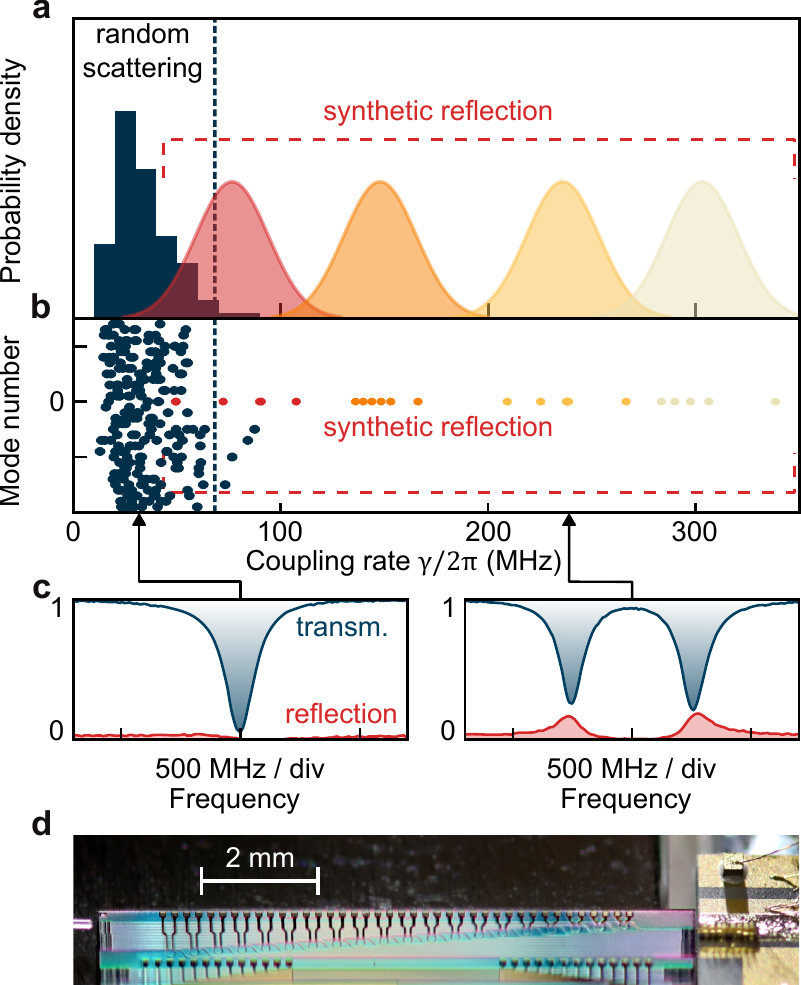}
  \caption{
    \highlightchanges{\textbf{Resonator characterization}.
    \textbf{a}, Distributions of measured forward-backward coupling rates $\gamma/(2\pi)$ in resonances with random backscattering,  unaffected by synthetic reflection (blue histogram) and those that are deliberately affected by the by the corrugation with four select corrugation amplitudes (red, orange, yellow and beige-shaded indicative Gaussian distributions).
    \textbf{b}, Scatter-plot, coupling rate vs. mode number of the data underlying (a). Synthetic reflection impacts only the pumped mode with relative mode number $\mu=0$. 
    \textbf{c}, Nonzero forward-backward coupling manifests itself as a splitting of transmission (blue) and reflection lineshapes. Left: For $\gamma < \kappa$ the splitting is unresolved; right: For $\gamma > \kappa$ the splitting is apparent.
    \textbf{d}, Photograph of the experimental system showing the semiconductor laser diode (right) butt-coupled to the photonic chip  with the PhCRs (middle). Transmitted light is out-coupled using an optical fiber (left).
    }}
  \label{fig:exp1}
\end{figure}

\highlightchanges{The presented criteria enable us to tailor the synthetic backreflection and to demonstrate a self-injection locked soliton source, operating deterministically and exclusively in the single-DKS regime by design.}

\begin{figure*}[t]
  \centering
  \includegraphics[width=\textwidth]{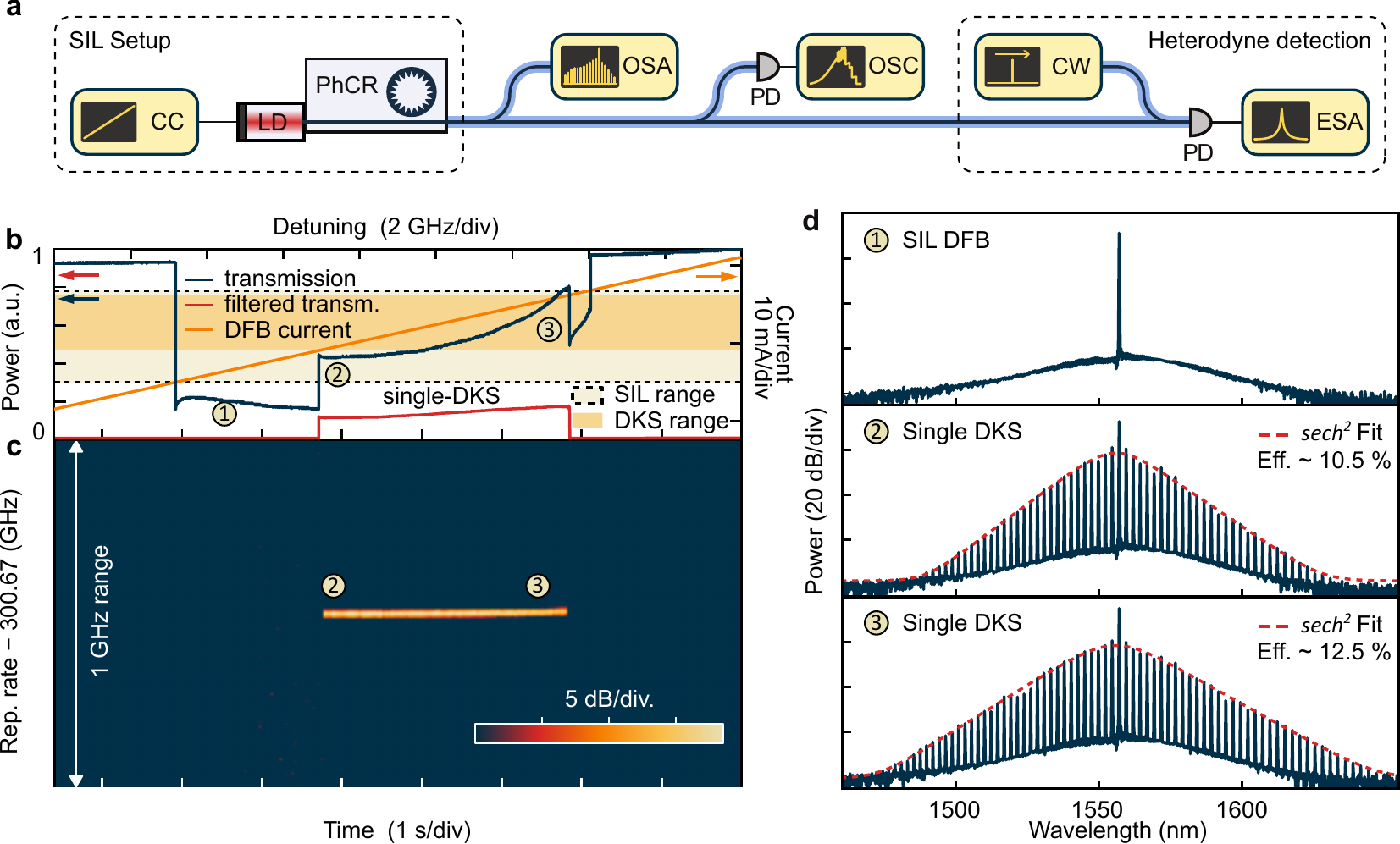}
  \caption{
    \highlightchanges{\textbf{Synthetic reflection enabled DKS generation}.
    \textbf{a}, Experimental setup. CC, current controller; LD, laser diode; OSA and ESA, optical and electrical spectrum analyzers, respectively; OSC, oscilloscope; CW, continuous-wave laser; PD, photodiode.
    \textbf{b} Total transmission (blue) and bandpass-filtered microcomb power (cf. main text) (red; filter offset from the pump, indicates comb formation) measured during a laser scan towards longer wavelengths. The orange line corresponds to the driving current. SIL and (single) DKS ranges on the current axis are highlighted. The horizontal axis indicates the free-running laser detuning and the experiment time. 
    \textbf{c}, Repetition rate signal recorded during the same laser scan as in (b), measured via heterodyne detection (cf. main text).    
    \textbf{d}, Optical spectra measured within the SIL range: CW SIL \Circled{1}; SIL-based single DKS states \Circled{2} and \Circled{3} at different detuning as indicated by the corresponding circled numbers in (b).}
    }
  \label{fig:exp2}
\end{figure*}

In preparation for the experiments, a range of \highlightchanges{photonic chip-integrated silicon nitride PhCRs (embedded in a silica cladding)} are fabricated with varying corrugation amplitude \highlightchanges{(ca. 10-50~nm)}, in a commercial foundry process (Ligentec) \highlightchanges{based on ultraviolet (UV) stepper lithography, compatible with wafer-level production.} 
The resonators' free-spectral range (FSR) is 300~GHz (radius~75~µm) \highlightchanges{and their waveguide's height and mean witdh of 800~nm and 1600~nm, are chosen to provide anomalous disperion as required for DKS formation.}  
We characterize the fabricated resonators \highlightchanges{by recording their transmission spectrum in a} frequency comb-calibrated laser scans \cite{delhaye:2009}, permitting to retrieve the coupling rates $\gamma$ and the resonance widths $\kappa$ \highlightchanges{via lineshape fitting} over a broad spectral bandwidth. 
\highlightchanges{\reffig{fig:exp1}{a} shows the retrieved distributions of the coupling rate $\gamma$ of multiple PhCRs with different corrugation amplitudes. The blue histogram reports the distribution of $\frac{\gamma}{2\pi}$ for the resonances that are not affected by the corrugation (i.e. away from the pumped mode) and the red, orange, yellow, and beige shadings represent the distributions of the deliberately split resonances for PhCR designs with different corrugation amplitudes (small to large, 6 samples for each amplitude). The underlying data is shown in \reffig{fig:exp1}{b} as a scatter plot.
Although the random imperfection-based forward-backward coupling rate $\frac{\gamma}{2\pi}$ can in rare cases reach significant values (here: up to $\sim 100$~MHz, corresponding to $2\gamma/\kappa\approx2$), the most probable value is low (here: $\sim 25$~MHz, corresponding to $2\gamma/\kappa\approx0.5$). In contrast, PhCRs enables control over the backscattering rate by design and, as shown by the experimental data in \reffig{fig:exp1}, provides robust access to backreflections that are difficult or impossible to access through random imperfections.
Importantly, only a single pre-defined resonance to which the PhCR's corrugation is matched, exhibits significant forward-backward coupling, while all other modes remain unaffected. This is evidenced by the data in \reffig{fig:exp1}{b} and further illustrated in the SI, Figure~S3a. \highlightchanges{Established concepts of waveguide dispersion engineering to create dispersive waves or broadband spectra remain unaffected.}
No noticeable degradation of the $Q$-factor is observed up to $\gamma/2\pi \lesssim$~5~GHz (approximately 100$\times$ above what is used here), corresponding to a critically coupled linewidth of $\frac{\kappa}{2\pi} \approx 110$~MHz (SI, Figure~S3).}

For the experiments, a semiconductor distributed feedback laser diode (DFB) is butt-coupled to the photonic chip \highlightchanges{(see \reffig{fig:exp1}{d})}, permitting an estimated on-chip pump power of \highlightchanges{$P=21$~mW}, corresponding to \highlightchanges{$f^2\approx7.3$}. From Eqs.~\ref{eq:f2th} and \ref{eq:gamma1DKS}, we obtain an ideal backscattering range of \highlightchanges{$2\gamma/\kappa \in (1.83, 3.35)$}, ensuring \highlightchanges{deterministic and exclusive generation of a single DKS.} 
Based on these considerations we choose a PhCR with a \highlightchanges{normalized synthetic backscattering for the pump mode at 1557~nm of $2\gamma/\kappa \approx 2.67$ ($\frac{\gamma}{2\pi} \approx$~145~MHz). This PhCR is critically coupled and exhibits anomalous group velocity dispersion ($D_2 \approx 9$~MHz; cf. Methods for a definition).} 
The DFB pump laser diode is mounted on a piezo translation stage to adjust the injection phase \cite{Kondratiev2017}, an actuator which can readily be achieved through on-chip heaters \cite{xiang:2021}; to reduce the device footprint and allow for more resonators on the chip, we have omitted this feature. \highlightchanges{The transmitted light is collected by an optical fiber whose tip is immersed in index-matching gel to suppress parasitic reflection.} An overview of the setup is shown in \reffig{fig:exp2}{a}. The laser's emission wavelength can be tuned via its drive current. As long as the laser diode does not receive a resonant injection from the microresonator it is free-running. When it is close in frequency to the microresonator resonance, a strong resonant backward wave is generated, providing frequency-selective optical feedback resulting in SIL. 

We slowly (within ca.~10~s) tune the DFB's electrical drive current to scan the emission wavelength across the pump resonance, towards longer wavelength. During the laser scan, we monitor the transmitted power as well as the power of a filtered spectral portion of the long-wavelength wing of the generated microcombs (as an indicator for comb formation). Both are shown along with the DFB diode's drive current in \reffig{fig:exp2}{b}. 
Here, the SIL regime is clearly evidenced by a sharp drop of the transmission that, after optimizing the injection phase, extends over a wide range of electrical drive current values. The DKS regime is marked by the non-zero filtered transmitted power. 


To confirm DKS generation, we record the 300~GHz DKS repetition rate beatnote and the optical spectrum. As the repetition rate signal is not directly detectable, modulation sidebands around a pair of adjacent DKS comb lines are generated electro-optically. Their heterodyne beating with the DKS comb lines creates a signal at lower frequency, from which the repetition rate can be reconstructed \cite{delhaye:2012}.
\reffig{fig:exp2}{c} shows the reconstructed repetition rate signal obtained during the DFB laser scan, and \reffig{fig:exp2}{d} shows optical spectra that correspond to different current values.

Upon entering the SIL regime, we observe at first only the single optical frequency of the SIL pump laser (\reffig{fig:exp2}{d}~\Circled{1}). Continuing the scan we next observe an abrupt transition into a single-DKS microcomb state (\reffig{fig:exp2}{d}~\Circled{2}). Such single-DKS states are characterized by a smooth squared hyperbolic-secant amplitude and a pulse repetition rate that corresponds to the resonator's FSR; these properties are highly-desirable for applications.
\highlightchanges{As the scan proceeds the spectrum becomes broader (\reffig{fig:exp2}{d}~\Circled{3}), which is an experimental manifestation of access to a range of detunings; this is also manifest in the slope of the transmission signal in the DKS state.}
Scanning even further causes the DKS to disappear, and the system to return to CW SIL operation (spectrum similar to \reffig{fig:exp2}{d}~\Circled{1}), before eventually exiting the SIL regime entirely. When repeated, each scan shows the same SIL dynamics, including deterministic single-DKS generation, independent of the scan speed (SI, Figure~S5). 
Turing patterns, noisy comb-states and multi-DKS regimes are absent in contrast to previously demonstrated self-injection locked DKS. 
\highlightchanges{A detailed comparison between samples with different levels of backscattering $2\gamma/\kappa$ is provided in the SI, Figure~S2, showing noisy or multi-DKS states as well as drastically reduced extend of the single-DKS range for low values of $2\gamma/\kappa$.}

Although not pursued here, we note that the pump to DKS conversion efficiency in the states \Circled{2} and \Circled{3} is 10.5~\% and 12.5~\%, resp., significantly higher than what would be expected in conventional resonators. This is a consequence of the mode splitting, shifting the pumped resonance effectively closer to the pump laser as explored previously in coupled ring-resonators \cite{helgason_power-efficient_2022}. \highlightchanges{We also confirm that synthetic reflection does not lead to higher noise in injection locked regime, neither for a continuous-wave lasing nor for a DKS as further detailed in the SI (Figure~S4).}

\begin{figure}[t]
  \centering
  \includegraphics[width=\columnwidth]{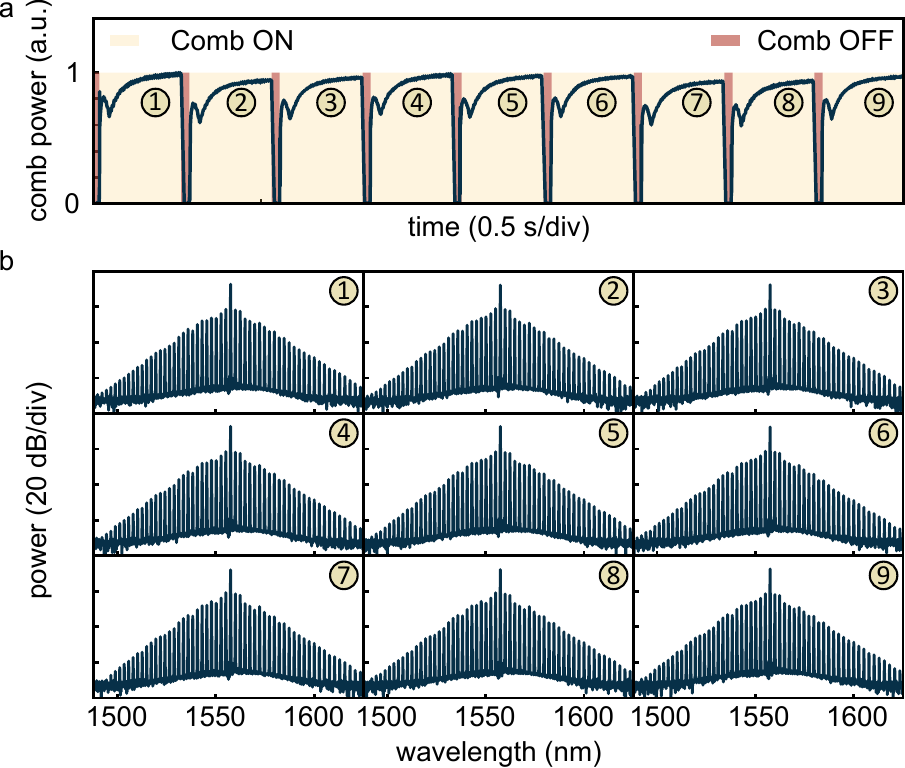}
  \caption{
    \highlightchanges{\textbf{Determinitic \textit{on-off} switching of single-DKS}.
    \textbf{a}, Bandpass-filtered power (filter is offset from pump) measured during 9 consecutive \textit{on}-\textit{off} switching cycles. In each cycle, the laser diode current is abruptly turned off and then ramped back up to the preset value.
    \textbf{b}, Optical spectra recorded in the respective \textit{on}-phase of each cycle.}
    }
  \label{fig:exp3}
\end{figure}

\highlightchanges{To demonstrate the deterministic and exclusive single-DKS generation, we repeatedly turn the diode's current \textit{on} and \textit{off} by an automated procedure.  We monitor the comb power (filtered transmission) (\reffig{fig:exp3}{a}) and record the optical spectrum created in each \textit{on-off} cycle (\reffig{fig:exp3}{b}). Each time a single DKS is generated.}

\subsection*{Conclusion}
In conclusion, we have demonstrated microresonators with synthetic reflection and achieve a self-injection locked soliton source, operating deterministically and exclusively in the desirable single-DKS regime by design. 
The presented results in conjunction with the scalable, widely accessible fabrication process, the low-cost components, and, notably, its ease of operation meet important requirements of out-of-lab applications. Further research may explore extending the presented results to combs in the backward direction (cf. SI), effectively blue-detuned DKS combs with potentially even higher conversion efficiency \cite{helgason_power-efficient_2022} or multiple pump wavelengths \cite{Chermoshentsev2022}. In addition, the novel concept of synthetic reflection may be transferred to other integrated photonic systems, including normal dispersion combs \cite{xue:2016, lihachev:2022}, \highlightchanges{optical parametric oscillators \cite{black_optical-parametric_2022}, integrated tunable lasers \cite{corato-zanarella:2023, lihachev:2023}} and novel quantum light sources \cite{zhao:2020, lu:2022}.

\subsection*{Methods}
\small

\paragraph{Numerical model.} 
To simulate the nonlinear DKS and breathing DKS existence range in Figure~\ref{fig:intro}c we consider a system of coupled mode equations \cite{chembo:2010, hansson:2014} for forward $a_\mu$ and backward $b_\mu$ mode amplitudes, where 
$\mu$ denotes the relative (longitudinal) mode number with respect to the pump mode ($m_0 \leftrightarrow \mu = 0$):
\begin{align*}
  \partial_t a_\mu = &-(1+i\zeta_\mu)a_\mu + i\sum_{\mu^\prime = \nu + \eta - \mu} a_\nu a_\eta a_{\mu^\prime}^* + 2i a_\mu \sum_{\eta} |b_\eta|^2 +\\ &+ i\delta_{\mu 0}\frac{2\gamma}{\kappa} b_\mu + f \delta_{\mu 0}\\ 
  \partial_t b_\mu = &-(1+i\zeta_\mu)b_\mu + i\sum_{\mu^\prime = \nu + \eta - \mu} b_\nu b_\eta b_{\mu^\prime}^* + 2i b_\mu \sum_{\eta} |a_\eta|^2 +\\ &+ i\delta_{\mu 0}\frac{2\gamma}{\kappa} a_\mu
\end{align*}
where $\zeta_\mu = \frac{2}{\kappa} (\omega_{\mu} - \omega_p - \mu D_1)$ is a dimensionless detuning defined by the pump laser frequency $\omega_p$ and the resonance frequencies $\omega_\mu = \omega_0 + D_1 \mu + \frac{1}{2}D_2 \mu^2$ (where $D_1/2 \pi$ and $D_2/2 \pi$ correspond to the FSR and the GVD, \highlightchanges{and $\omega_0$ is the resonance frequency of the pumped mode); $f =\sqrt{8\eta \omega_0c n_2 P / (\kappa^2 n^2 V_\mathrm{eff})}$ is the normalized pump power, with the coupling coefficient $\eta=1/2$ (critical coupling), $c$ the speed of light, $P$ the pump power, $n$ the refractive index, $n_2$ the nonlinear refractive index and $V_\mathrm{eff}$ the effective mode volume; note that $f_\mathrm{th}$ in Eq.~\ref{eq:f2th} is normalized the same way as $f$}. The third term in each equation corresponds to the cross-phase modulation by the respective counter-propagating waves, while the fourth term represents the coupling between forward and backward propagating waves. Instead of modeling the SIL dynamics by including laser rate equations, we numerically define the detuning. This approach cannot describe the abrupt transition from the free-running laser to the SIL state, it remains however valid for the specified detuning and can qualitatively capture the features observed in the experiment. Simulation parameters similar to those of the experimental system are used. The numerical simulation also enables us to compute the nonlinear dispersion as illustrated in Figure~S1, SI.


\subsection*{Funding}
\small
This project has received funding from the European Research Council (ERC) under the EU’s Horizon 2020 research and innovation program (grant agreement No 853564), from the EU’s Horizon 2020 research and innovation program (grant agreement No 965124) and through the Helmholtz Young Investigators Group VH-NG-1404; the work was supported through the Maxwell computational resources operated at DESY.

\subsection*{Data availability}
\small The datasets generated and analysed during the current study are available from the corresponding author on reasonable request.

\subsection*{Code availability}
\small Numeric simulation codes used in the current study are available from the corresponding author on reasonable request.

\subsection*{Competing interests}
\small We declare that none of the authors have competing interests; J.D.J. and M.K. are cofounders of Enlightra.




\printbibliography
\end{refsection}

\title{Synthetic-self-injection locked microcombs - \\ Supplemental Information}

\newpage
\onecolumn

\maketitle

\setcounter{equation}{0}
\setcounter{figure}{0}
\renewcommand{\thefigure}{S\arabic{figure}}

\begin{refsection}
\section{Coupled mode equations and pump mode hybridization}
\label{sec:CME}
The normalized coupled mode equations (CMEs) for the pump in forward and backward directions read
\begin{align}
        \partial_t a_0 & = -(1+i\zeta_0)a_0 + i|a_0|^2a_0 + 2i|b_0|^2a_0 + i\beta b_0 +f 
        \label{eq:cme_a}\\
        \partial_t b_0 & = -(1+i\zeta_0)b_0 + i|b_0|^2b_0 + 2i|a_0|^2b_0 + i\beta a_0  \label{eq:cme_b}
\end{align}
where for convenience the normalized coupling rate $\beta=2\gamma/\kappa\geq0$ has been introduced. Without loss of generality, $f\geq 0$. \\
\\
\noindent The coefficient matrix of the  system of equations  (without the pump) 
\begin{align}
M=
    \begin{pmatrix}
        -(1+i\zeta_0) + i|a_0|^2 + 2i|b_0|^2 & i\beta \\
        i\beta & -(1+i\zeta_0) + i|b_0|^2 + 2i|a_0|^2
    \end{pmatrix}
\end{align}
has the following Eigenvalues
\begin{align}
    \lambda_\pm & = - \left( 1 + i\zeta_0 -\frac{3}{2}i (|a_0|^2 + |b_0|^2) \pm i\sqrt{\beta^2 + \left(\frac{1}{2}(|a_0|^2 - |b_0|^2)\right)^2}\right) \\
    & = - \left( 1 + i(\zeta_0 -\delta \zeta_\mathrm{NL} \pm \sqrt{\beta^2 + \delta \beta_\mathrm{NL}^2})\right)
\end{align}
and is diagonalized in the following Eigenbasis of hybridized forward-backward modes
\begin{align}
    & \left\{ \frac{1}{2}(|a_0|^2 - |b_0|^2) \pm \sqrt{\beta^2 + \left(\frac{1}{2}(|a_0|^2 - |b_0|^2)\right)^2}; \, - \beta  \right\}\\
    & =  \left\{\delta \beta_\mathrm{NL} \pm \sqrt{\beta^2 + \delta \beta_\mathrm{NL}^2}; \, -\beta  \right\}
\end{align}
where $\delta\beta_\mathrm{NL} = \frac{1}{2}(|a_0|^2 - |b_0|^2)$ and $\delta\zeta_\mathrm{NL}=\frac{3}{2}(|a_0|^2 + |b_0|^2)$.
The transformation matrices are:
\begin{align}
T=
    \begin{pmatrix}
        \delta \beta_\mathrm{NL} + \sqrt{\beta^2 + \delta \beta_\mathrm{NL}^2}  & \delta \beta_\mathrm{NL} - \sqrt{\beta^2 + \delta \beta_\mathrm{NL}^2}  \\
    -\beta & -\beta
    \end{pmatrix}
\end{align}
and
\begin{align}
T^{-1}=
    \begin{pmatrix}
        \frac{1}{2\sqrt{\beta^2 + \delta \beta_\mathrm{NL}^2}}   & \frac{\delta\beta_\mathrm{NL} - \sqrt{\beta^2 + \delta \beta_\mathrm{NL}^2}}{2\beta\sqrt{\beta^2 + \delta \beta_\mathrm{NL}^2}}   \\
    -\frac{1}{2\sqrt{\beta^2 + \delta \beta_\mathrm{NL}^2}} & \frac{-\delta\beta_\mathrm{NL} - \sqrt{\beta^2 + \delta \beta_\mathrm{NL}^2}}{2\beta\sqrt{\beta^2 + \delta \beta_\mathrm{NL}^2}}
    \end{pmatrix}
\end{align}
so that 
\begin{align}
    \begin{pmatrix}
        a_0 \\ b_0
    \end{pmatrix}
    = T
    \begin{pmatrix}
        a_+ \\ a_-
    \end{pmatrix}
\end{align}
and
\begin{align}
    \begin{pmatrix}
        a_+ \\ a_-
    \end{pmatrix}
    = T^{-1}
    \begin{pmatrix}
        a_0 \\ b_0
    \end{pmatrix}
\end{align}
where $a_\pm$ denote the (not specifically normalized) field amplitudes of the hybrid modes. The steady state equations for the hybrid modes are
\begin{align}
    0 & = -\left( 1 + i(\zeta_0 -\delta \zeta_\mathrm{NL} + \sqrt{\beta^2 + \delta \beta_\mathrm{NL}^2})\right) a_+ + \frac{1}{2\sqrt{\beta^2 + \delta \beta_\mathrm{NL}^2}} f \\
    0 & = -\left( 1 + i(\zeta_0 -\delta \zeta_\mathrm{NL} - \sqrt{\beta^2 + \delta \beta_\mathrm{NL}^2})\right) a_- - \frac{1}{2\sqrt{\beta^2 + \delta \beta_\mathrm{NL}^2}} f
\end{align}
In non-normalized units, the effective resonance frequencies of the hybridized modes are
\begin{align}
    \omega_\mathrm{\pm, eff} = \omega_0 - \frac{\kappa}{2}\left( \delta\zeta_\mathrm{NL}  \pm \sqrt{\beta^2 + \delta \beta_\mathrm{NL}^2} \right)
\end{align}

\section{Approximations for the forward pump mode under strong coupling}
In what follows, it is assumed that
\begin{itemize}
    \item the coupling is strong $\beta>1$
    \item due to the strong coupling, the power levels in forward and backward directions are approximately equal $|a_0|^2 = |b_0|^2$. Note that due to symmetry breaking \cite{delbino:2017} this is only valid up to a certain power level. We validated, by numeric integration of the CMEs~\ref{eq:cme_a} and \ref{eq:cme_b}, that this approximation is valid.
    \item the detuning $\zeta_0$ is such that approximately only the lower frequency hybrid mode $a_-$ is driven, i.e. $|a_-|\gg|a_+|$.
\end{itemize}
Under these assumptions,
\begin{align}
    a_0 & = \left(\delta \beta_\mathrm{NL} - \sqrt{\beta^2 + \delta \beta_\mathrm{NL}^2}\right) a_- \approx -\beta a_-\\
    b_0 & = - \beta a_- 
\end{align}
and in consequence
\begin{align}
    (1+i(\zeta_0 - \delta\zeta_\mathrm{NL} -\beta))a_0 = \frac{f}{2}
\end{align}
Multiplying each side of the equation with its complex conjugate results in
\begin{align}
    (1+(\zeta_0 - \delta\zeta_\mathrm{NL} -\beta)^2)|a_0|^2 = \frac{f^2}{4} 
\end{align}
An immediate insight is that the strong coupling between forward and backward waves limits the power in the forward (or backward) wave to values of 
\begin{align}
    |a_0|^2 \leq f^2/4
    \label{eq:max_P}
\end{align}
Expressing $\delta\zeta_\mathrm{NL}$ via the field amplitudes gives
\begin{align}
    (1+(\zeta_0 - 3|a_0|^2 -\beta)^2)|a_0|^2 = \frac{f^2}{4} 
\end{align}
and for the detuning
\begin{align}
    \zeta_{0,\pm} = \beta + 3|a_0|^2 \pm \sqrt{\frac{f^2}{4|a_0|^2}-1}
    \label{eq:detuning}
\end{align}
where $\zeta_{0,+}$ corresponds to an effective red-detuning and $\zeta_{0,-}$ to an effective blue-detuning with regard to the lower-frequency hybrid mode $a_-$.

\section{Threshold condition and first oscillating sideband}
We consider two initially zero-power (except for vaccuum fluctuations) sidebands with mode number $\pm\mu$ relative to the pumped mode. Their CMEs are
\begin{align}
    \partial_t a_{+\mu} & = -(1+i(\zeta_{\mu} - 4|a_0|^2)) a_{+\mu} + ia_0^2 a_{-\mu}^* \\
    \partial_t a_{-\mu}^* & = -(1-i(\zeta_{\mu} - 4|a_0|^2)) a_{-\mu}^* - ia_0^{*2} a_{+\mu}
\end{align}
where again $|a_0|^2 \approx |b_0|^2$ was assumed and $\zeta_{\mu}=\frac{2}{\kappa}(\omega_0 - \omega_\mathrm{p} + \frac{1}{2}D_2\mu^2)=\zeta_0 + \frac{D_2}{\kappa}\mu^2$. The Eigenvalues of this set of equations are
\begin{align}
    \lambda_\pm = -1 \pm \sqrt{|a_0|^4 - (\zeta_\mu - 4|a_0|^2)^2}
\end{align}
The parametric gain experienced by the two sidebands therefore is
\begin{align}
    G = \kappa\sqrt{|a_0|^4 - \left(\zeta_0 +\frac{D_2}{\kappa}\mu^2 - 4|a_0|^2\right)^2}
    \label{eq:paramgain}
\end{align}
At least a intracavity power of $|a_0|^2=1$ is required to reach threshold. With Eq.~\ref{eq:max_P} it follows that for strong coupling the threshold pump power $f^2 \geq 4|a_0|$ is at least four times the threshold power of a resonator without forward-backward coupling.

\subsection{First oscillating sideband \highlightchanges{(Condition for single-DKS)}} \highlightchanges{We are interested in finding the condition under which the first oscillating modulation instability (MI) sidebands will appear in the resonances directly adjacent to the pump mode; as described in the main text, this would seed a single DKS pulse.}

The phase mismatch between the pump wave and the resonator modes can be quantified via their effective (including nonlinear frequency shifts) detuning $\zeta_\mathrm{\mu, eff}$ from an equidistant $D_1$-space frequency grid. A smaller $\zeta_\mathrm{\mu, eff}$ implies better phase matching. 
\begin{align}
    \zeta_\mathrm{\mu, eff} & = \zeta_0 + \frac{D_2}{\kappa}\mu^2 - 4|a_0|^2 \\
    & = \beta - |a_0|^2  \pm \sqrt{\frac{f^2}{4|a_0|^2}-1} + \frac{D_2}{\kappa}\mu^2 
\end{align}

 For DKS the resonator is characterized by anomalous dispersion $D_2>0$ ($\beta \gg D_2/\kappa $). It can therefore be guaranteed, that the first generated sideband pair (best phase matching) will be $\mu\pm1$, if 
 \begin{align}
     \beta - |a_0|^2 - \sqrt{\frac{f^2}{4|a_0|^2}-1} > 0
 \end{align}
Assuming $|a_0|^2 \leq f^2/4$ we find
  \begin{align}
     \boxed{\beta  > f^2/4 \quad \Leftrightarrow \quad  \gamma/\kappa  > f^2/8 } 
 \end{align}
 as a condition that guarantees that the first sideband pair will be generated at $\mu=\pm1$.

\subsection{Threshold power}
The threshold power $f_\mathrm{th}$ is the power level where the parametric threshold $G>\kappa$ can be reached. Inserting Eq.~\ref{eq:detuning} for the detuning into Eq.~\ref{eq:paramgain}, we obtain for the threshold condition
\begin{align}
& |a_0|^4 - \left(\beta + 3|a_0|^2 \pm \sqrt{\frac{f_\mathrm{th}^2}{4|a_0|^2}-1} +\frac{D_2}{\kappa}\mu^2 - 4|a_0|^2\right)^2 = 1 \\
    \Leftrightarrow \, & |a_0|^4 - 1 = \left(\beta  \pm \sqrt{\frac{f_\mathrm{th}^2}{4|a_0|^2}-1} +\frac{D_2}{\kappa}\mu^2 - |a_0|^2\right)^2 
\end{align}
Under the assumption that $\beta - |a_0|^2 - \sqrt{\frac{f_\mathrm{th}^2}{4|a_0|^2}-1} > 0$ (condition for first oscillating sidebands $\mu=\pm1$), this results in

\begin{align}
\boxed{f_\mathrm{th}^2  = 4|a_0|^2 + 4|a_0|^2 \left( \beta - |a_0|^2 +\frac{D_2}{\kappa}\mu^2  - \sqrt{|a_0|^4 - 1} \right)^2, \quad (|a_0|^4 > 1)}
\label{eq:th1}
\end{align}

This equation can be solved numerically. For example, $\beta=4$ results in $f_\mathrm{th}^2 \approx 8.4$

\subsection{Threshold power assuming zero-effective detuning}
A simplified threshold condition may be derived assuming that the threshold will be reached at zero effective detuning so that
\begin{align}
    \zeta_0 & = \beta + 3|a_0|^2 \, \hspace{0.5cm} \mathrm{and}\\
    f_\mathrm{th}^2 & = 4|a_0|^2
\end{align}
In this case the threshold condition is
\begin{align}
    & |a_0|^4 - \left(\beta + 3|a_0|^2 +\frac{D_2}{\kappa}\mu^2 - 4|a_0|^2\right)^2 = 1 \\
    \Leftrightarrow\,  & \frac{f_\mathrm{th}^4}{16} - \left(\beta - \frac{1}{4}f_\mathrm{th}^2 +\frac{D_2}{\kappa}\mu^2\right)^2 = 1 \\
    \Leftrightarrow\,  & f_\mathrm{th}^2 = \frac{2+2\left(\beta + \frac{D_2}{\kappa}\mu^2\right)^2}{\beta + \frac{D_2}{\kappa}\mu^2}  
\end{align}
Assuming that $\beta \gg \frac{D_2}{\kappa}\mu^2$ this simplifies to
\begin{align}
    \boxed{ f_\mathrm{th}^2 \approx \frac{2+2\beta^2}{\beta} = 2\beta + \frac{2}{\beta} \quad \Leftrightarrow \quad f_\mathrm{th}^2 \approx 4\frac{\gamma}{\kappa} + \frac{\kappa}{\gamma}}
\end{align}
For $\beta=4$ we find $f_\mathrm{th}^2 = 8.5$,  almost equal to what is obtained through Eq.~\ref{eq:th1}.

\section{Pump mode hybridization above threshold}

\begin{figure*}[h]
  \centering
  \includegraphics[width=400 pt]{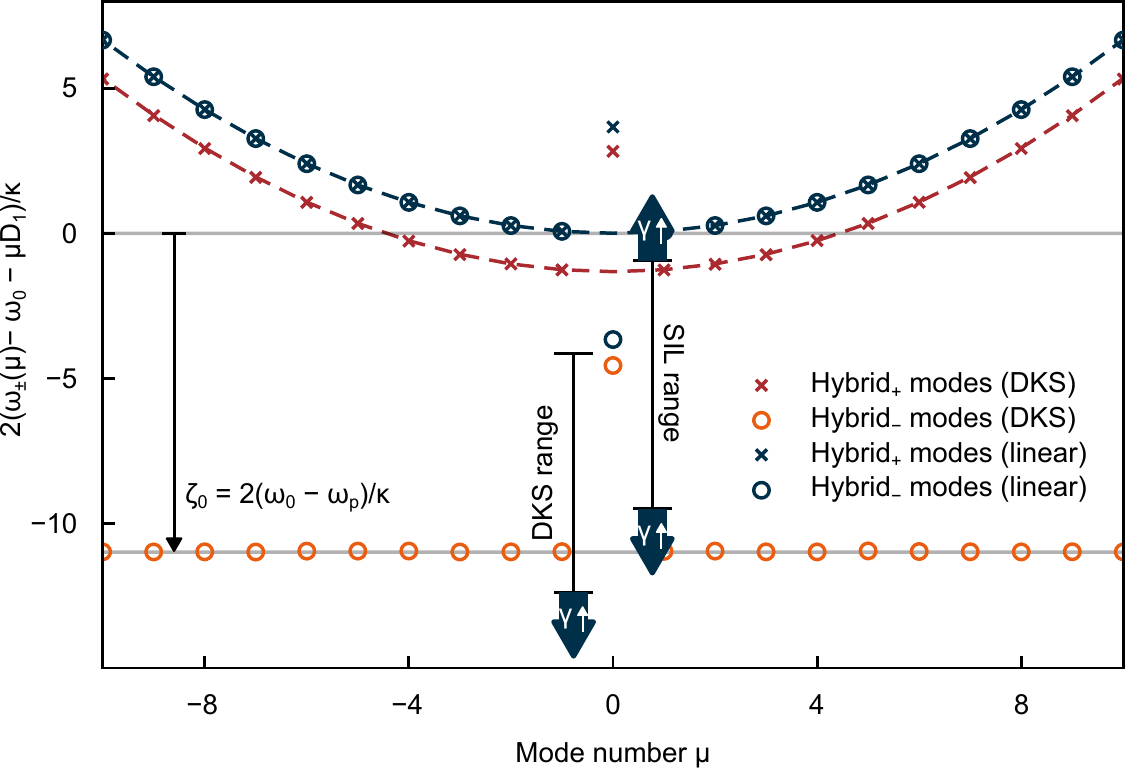}
  \caption{\highlightchanges{\textbf{Nonlinear integrated dispersion of the hybridized modes. } Linear (blue circles and crosses) and nonlinear (orange circles and red crosses) integrated dispersions of a PhCR as detailed in the text.
    }}
    \label{fig:nonlin_modes}
\end{figure*}

The derivations of Section~\ref{sec:CME} are only valid below threshold, where only the forward and backward pump mode are excited. Above threshold, and in particular in presence of DKS, the effective frequencies of the hybrid modes resulting from the avoided mode crossing of the coupled forward and backward modes are

\begin{align}
\omega_\mathrm{\pm, eff}(\mu) = \frac{1}{2}(\omega_\mathrm{a, eff}(\mu) + \omega_\mathrm{b, eff}(\mu)) \pm \frac{1}{2} \sqrt{(\omega_\mathrm{a, eff}(\mu) - \omega_\mathrm{b, eff}(\mu))^2 + 4\gamma(\mu)^2}
\label{eq:nonlin_freqs}
\end{align}
where $\omega_\mathrm{a, eff}$ and $\omega_\mathrm{b, eff}$ are the effective (i.e. taking nonlinear frequency shifts into account) resonance frequencies of the forward and backward modes, respectively; $\gamma(\mu)$ denotes the mode dependent coupling.
\begin{align}
    \omega_\mathrm{a, eff}(\mu) & = \omega_0 - \frac{\kappa}{2}\left(\operatorname{Re}(\hat{\mathcal{F}} [ |\psi_a (\theta)|^2 \psi_a (\theta) ]_\mu / a_\mu ) + 2\sum_{\eta} |b_\eta|^2 \right) \\
    \omega_\mathrm{b, eff}(\mu) & = \omega_0 - \frac{\kappa}{2}\left(\operatorname{Re}(\hat{\mathcal{F}} [ |\psi_b (\theta)|^2 \psi_b (\theta) ]_\mu / b_\mu ) + 2\sum_{\eta} |a_\eta|^2 \right) \\
\end{align}
where $\psi_{a} (\theta) = \hat{\mathcal{F}}^{-1}[a_\mu] $ and $\psi_{b} (\theta) = \hat{\mathcal{F}}^{-1}[b_\mu] $ are the spatio-temporal field profiles and $\hat{\mathcal{F}} [~.~]_\mu$ stands for the component corresponding to the $\mu$-th mode ($\hat{\mathcal{F}}$ denotes the Fourier transform). \highlightchanges{Thus the nonlinear integrated dispersion for the hybrid modes can be defined as follows
\begin{align}
\mathrm{D}^\mathrm{int}_\pm(\mu) = 2\left(\omega_\mathrm{\pm, eff}(\mu) - \omega_0 - \mu D_1\right)/\kappa
\label{eq:nonlin_disp}
\end{align}
\newpage
An example of the nonlinear integrated dispersion is shown in \reffig{fig:nonlin_modes}{} for a PhCR in a single DKS state with $f=3$, and $2\gamma/\kappa\approx3.7$ for the pumped mode $\mu=0$ (zero-forward backward coupling is assumed for all other modes). In addition, the DKS existence range (DKS range) and the range of detunings accessible via SIL (SIL range) are indicated. Both, DKS and SIL ranges, are obtained numerically (cf. main text). For comparison, also the linear integrated dispersion (i.e. in the absence of nonlinear mode shifts) is shown. 
}

\section{Comb generation in the backward direction}
Due to the initially similar power-levels in forward and backward pump modes, combs may in principle not only be generated in the forward, but also in the backward direction. Indeed, when the pump laser detuning is between the (effective) resonance frequencies of the hybridized modes, the backward-wave is usually stronger (despite the forward-pumping). For the parameters considered in this work, we found that this range of detuning does not overlap with the soliton existence range and backward combs were not observed. However, backward combs represent an interesting opportunity for additional research. Aside from backward comb generation we note, that backward modulation instability, can trigger forward modulation instability (and comb generation) and vice versa, through a non-zero forward-backward coupling of the modulation instability sidebands. This can readily be included in the numeric model by introducing non-zero $\gamma$ also for modes with $\mu\neq0$.

\highlightchanges{\section{SIL DKS generation for different backscattering strengths}}

\begin{figure*}[h]
  \centering
  \includegraphics[width=\textwidth]{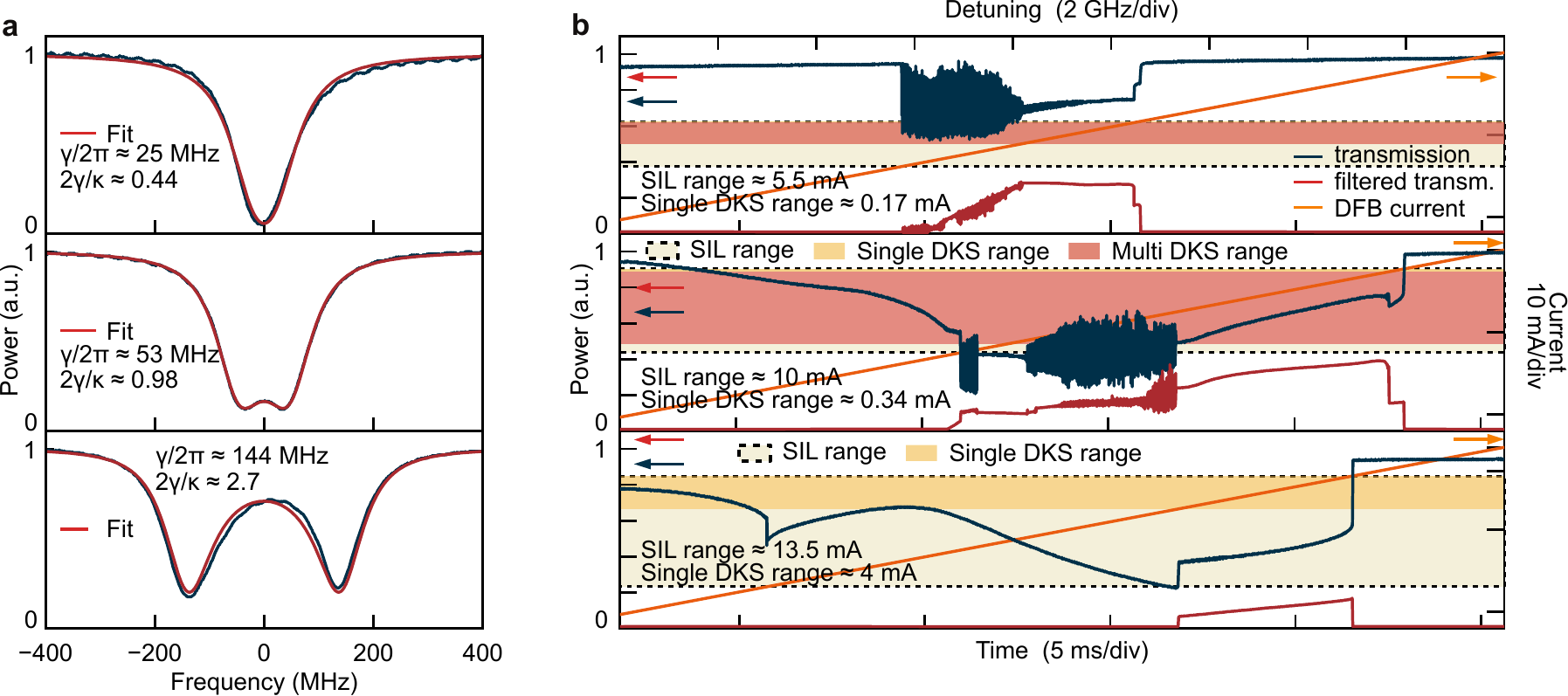}
  \caption{
    \highlightchanges{\textbf{SIL DKS generation for different backscattering strengths.}
    \textbf{a}, Measured lineshapes of microresonators with different level of backscattering (cf. main text for details). The linewidths $\kappa$ and coupling rates $\gamma$ (mode splittings) are retrieved from the fit.
    \textbf{b}, Total transmission (blue) and bandpass-filtered microcomb power (cf. main text) (red; filter offset from the pump, indicates comb formation) measured during a laser scan towards longer wavelengths. The orange line corresponds to the driving current. SIL range, as well as single- and multi-DKS ranges on the current axis are highlighted. The horizontal axis indicates the free-running laser detuning and the experiment time. 
    }    
    }
    \label{fig:s1}
\end{figure*}

\highlightchanges{We investigate experimentally SIL-based DKS generation in samples with different values of backscattering. The first sample has no corrugation pattern, but exhibits a strong random backscattering of $2\gamma/\kappa \approx 0.44$. The two remaining samples have a synthetic reflection of $2\gamma/\kappa \approx 0.98$ and 2.7.
The characterization of the pumped resonances is shown in \reffig{fig:s1}{a}. Total linewidths $\kappa$ and coupling rate $\gamma$ are extracted via lineshape fitting. 
Similar to Figure~3 in the main text, we explore self-injeciton locked DKS generation in these samples. The results are shown in 
\reffig{fig:s1}{b}. Provided a certain minimal back-scattering is present, all samples support a single-DKS state. However, the current range over which the single-DKS is observed is negligible for the weakest back-reflection and remains marginal for the medium level of back-reflection (compared to the current range over which multi-DKS and noisy comb states are generated). Only for the strongest back-reflection, a significant, readily accessible current range is achieved for the, in this case, exclusive single-DKS state. 
Note that we validated the single and multi-DKS regimes via their characteristic optical spectra (not shown).} 

\highlightchanges{\section{Characterization of PhCRs}}

\begin{figure*}[hb!]
  \centering
  \includegraphics[width=\textwidth]{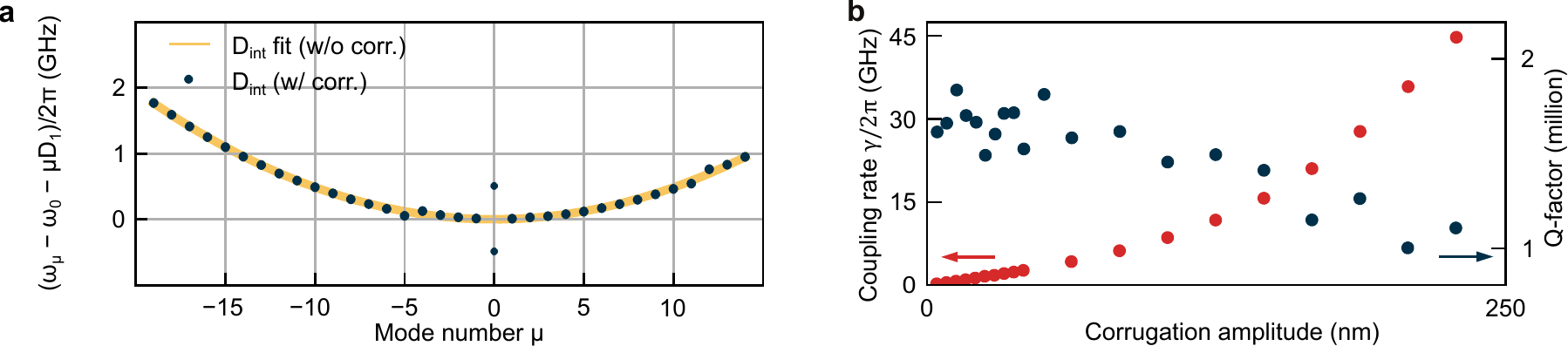}
   \caption{\highlightchanges{
    \textbf{PhCR characterization.}
    \textbf{a}, Integrated dispersion $D_\mathrm{int}$ for two samples: without corrugation pattern (yellow line), and with single-period corrugation (blue dots).
    \textbf{b}, Measured backward coupling rates $\gamma$ (red, left axis) and Q-factors (blue, right axis) of the PhCRs as a function of the estimated corrugation amplitude.
    }}
    \label{fig:s2}
\end{figure*}

\highlightchanges{All samples are characterized via frequency comb calibrated laser scans \cite{delhaye:2009}, which allow us to determine the resonator's dispersion, the coupling rates $\gamma$, the resonance widths $\kappa$, over a wide spectral bandwidth. The integrated dispersion $D_\mathrm{int} = (\omega_\mu-\omega_0 - \mu D_1)/2 \pi$ for a representative sample (with relatively large backscattering (for visibility) of $2\gamma/\kappa\approx10$) is shown in \reffig{fig:s2}{a} along with a conventional sample without corrugation. The comparison between both shows that, while the corrugation leads to a resonance splitting for the mode $\mu=0$ is has no effect on the overall dispersion profile.
\reffig{fig:s2}{b} shows the dependence of $\gamma$ and the $Q$-factor ($Q=\omega_0 /\kappa$) on the corrugation amplitude. No noticeable degradation of the $Q$-factor is observed up to $\gamma/2\pi \lesssim$~5~GHz; even for large coupling $\gamma\approx 45$~GHz, the Q-factor is only halved.}

\section{\highlightchanges{Noise properties of synthetic reflection SIL}}

\begin{figure*}[h]
  \centering
  \includegraphics[width=\textwidth]{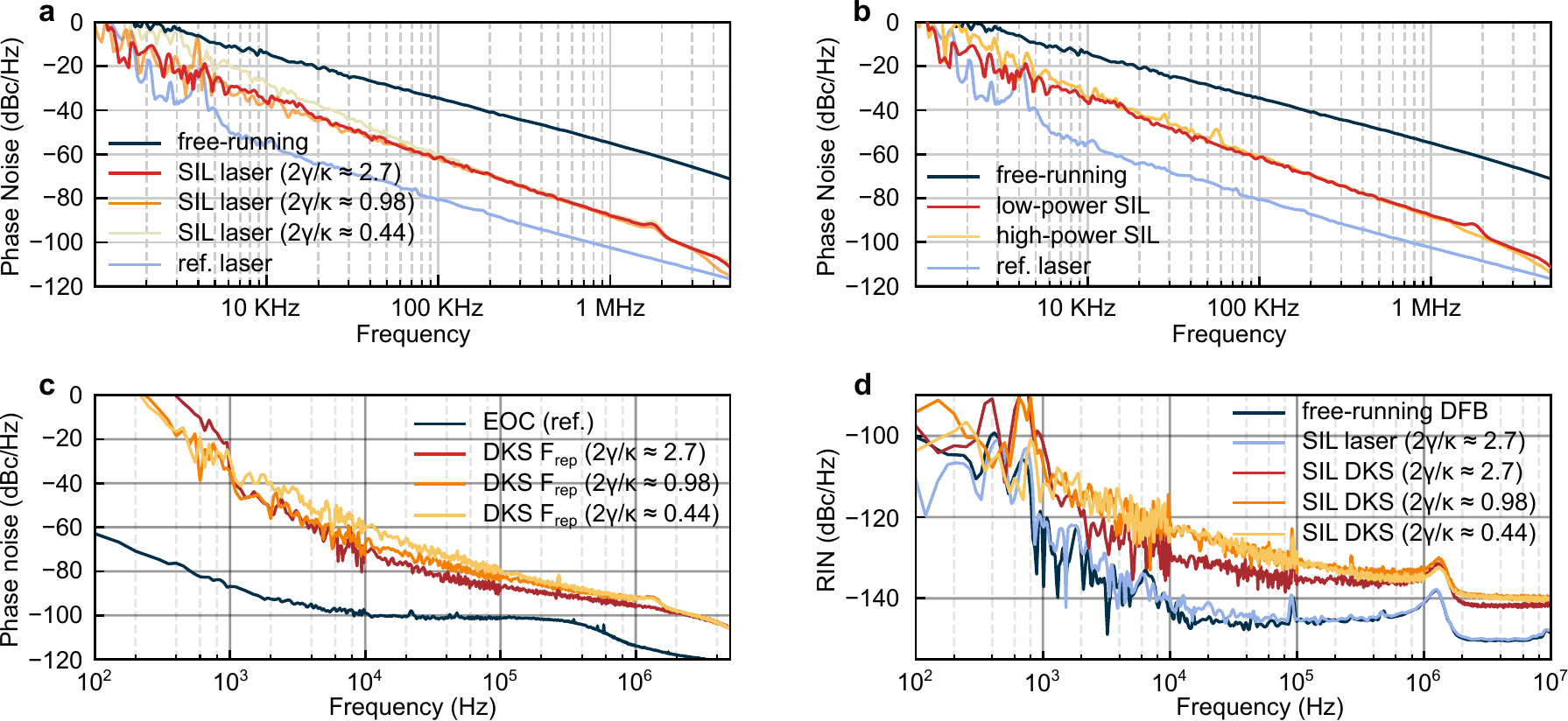}
   \caption{\highlightchanges{
    \textbf{Noise properties of SIL}.
    \textbf{a}, Phase noise of the DFB laser in free-running and SIL regimes  measured through heterodyne detection with a reference laser (cf. Figure~3, main text). The phase noise of the reference laser is provided as a baseline.
    \textbf{b}, Comparison of the phase noise of the DFB laser in SIL regimes at low and high output powers for the sample with $2\gamma/\kappa\approx2.7$
    \textbf{c}, Phase noise of the single DKS repetition rate. The phase noise of the electro-optic comb line used for heterodyning (cf. main text) is indicated for reference.
    \textbf{d}, Relative intensity noise (RIN) of SIL-based single DKS (red, ornage, yellow), low-power continuous-wave SIL lasing (blue) and the free-running DFB laser (dark blue).}    
    }
    \label{fig:noise}
\end{figure*}

\highlightchanges{\reffig{fig:noise}{} shows and compares noise properties continuous-wave lasing and DKS combs generated in samples with different levels of backscattering. \reffig{fig:noise}{a} shows the phase noise of the self-injection locked continuous-wave laser at low power and without soliton formation for different values of backscattering (the phase noise of the free-running laser diode as well as the phase noise of the reference laser are also shown). 
\reffig{fig:noise}{b} compares the phase noise of the self-injection locked continuous-wave laser at low power and high power (note that at high power a DKS is present). 
\reffig{fig:noise}{c} shows the repetition rate phase noise, i.e. the phase noise of the 300 GHz tone. Reduced noise is observed for larger values of backscattering. As the 300 GHz signal cannot be detected, an electro-optic comb (EOC) is used for optical-down mixing, as described in the main text (the EOC's phase noise is given for reference). 
\reffig{fig:noise}{d} compares the RINs of the free-running laser, the self-injection locked continuous-wave laser at low power, as well as the RIN of the DKS states.}

\highlightchanges{
\section{Comparison of different diode current ramp speeds}

\begin{figure*}[hb!]
  \centering
  \includegraphics[width=\textwidth]{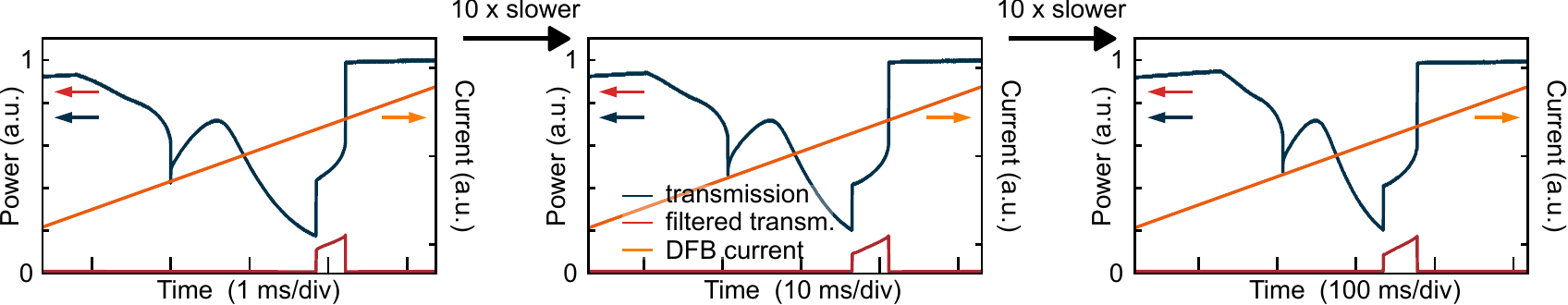}
  \caption{\highlightchanges{
    \textbf{SIL DKS formation at different diode current ramp speeds}.
    Filtered and unfiltered transmission recorded while ramping the diode current with different speed over the same current interval. The observed transmission (filtered/unfiltered) is virtually independent of the diode current ramp speed.}
    }
    \label{fig:scan_speed}
\end{figure*}

}

\printbibliography
\end{refsection}

\end{document}